\documentclass{article}
\usepackage{amsmath,amssymb}
\usepackage[noblocks]{authblk}
\usepackage{epstopdf}
\usepackage{cite}
\usepackage{siunitx}
\usepackage{color}
\usepackage{ulem}
\usepackage{here}
\usepackage{appendix}
\usepackage{graphicx}
\usepackage{wrapfig}
\usepackage{bm}
\usepackage{comment}
\usepackage{sidecap}
\usepackage[top=28truemm,bottom=28truemm]{geometry}

\begin{document}
\title{
The synchronization of elevators when not all passengers will ride the first-arriving elevator
}
\author[1]{Sakurako Tanida}
\affil[1]{Research Center for Advanced Science and Technology The University of Tokyo, 4-6-1 Komaba, Meguro-ku, Tokyo, Japan}
\affil[ ]{\textit {u-tanida@g.ecc.u-tokyo.ac.jp}}
\date{\today}
\maketitle

\begin{abstract}
The cluster motion of busy multiple elevators is considered to be one of the synchronization phenomena of autonomous oscillators. In this study, we used numerical simulations to quantitatively explore differences in the elevators' dynamics during down peaks under isolated and coupled conditions. We introduced the proportion of passengers who are set to ride the first-arriving elevator as a control parameter and investigated the behaviors of elevators when the proportions of those passengers and the inflow of passengers were varied. When we increased the inflow of passengers, the synchronization was promoted and the round-trip time increased. On the other hand, when we increased the proportion of those passengers, the synchronization was promoted while the round-trip time decreased. To elucidate the relationship between the parameters and dynamics and clarify the mechanism, we established simple mathematical models; First, we reproduced the round-trip time by a self-consistent equation considering the inflow of passengers and the proportion of passengers set to ride the first-arriving elevator. Then, we estimated an order parameter of the synchronization.
\end{abstract}
%
%
\maketitle

\section{Introduction}
Dynamics and order formation in nonequilibrium systems have garnered widespread interest in physics. Nonequilibrium systems are characterized by having an exchange of energy, particles, and/or information with their environment. Traffic dynamics exhibit nontrivial nonequilibrium nonlinear phenomena~\cite{Nagatani2002a,Helbing2001,Chowdhury2000b}. For example, in a traffic flow, the number of vehicles passing at a point per unit time increases as the vehicle density increases, whereas it fluctuates greatly and shows a metastable state when the density exceeds a critical value~\cite{Hall1986,Neubert1999}. Another example is the density wave emerging at a high density~\cite{Musha1976,Musha1978,Kerner1993,Komatsu1995,Kurtze1995}. Regarding the traffic flow as a nonequilibrium nonlinear system leads to countermeasures for the social problem of traffic congestion. In particular, the decelerating interaction when the distance between vehicles is short decreases the average speed of vehicles and causes traffic jams. Thus, congestion reduction solutions that remove deceleration factors other than safety measures have been successful.

Under the conditions of urban population concentration and increasing high-rise buildings, elevators are essential in the movement of people and exhibit interesting characteristic behaviors. For example, for a small number of users, elevators move as a one-to-one correspondence to the call by each user. As the number of passengers increases, multiple calls are responded to by one round-trip, and the elevator movement becomes oscillating gradually. More interestingly, multiple elevators sometimes spontaneously move closer to each other and move together~\cite{Poschel1994,Hikihara1997,Nagatani2002,Nagatani2003,Nagatani2004,Nagatani2011,Nagatani2012,Nagatani2015,Nagatani2016,Feng2021}. Such cluster motion appears to be similar to traffic jams; however, while traffic jams are caused by increasing vehicle densities, increasing the number of elevators will alleviate cluster motion. The factor inducing cluster motions is to increase the number of passengers. Therefore, the cluster motion is caused by a mechanism different from traffic jams. Recent studies suggest that elevator cluster motion is a type of oscillator synchronization phenomenon.

Synchronization is the adjustment of rhythms of autonomous oscillators using their weak coupling interactions~\cite{Pikovsky2001}. Synchronization phenomena are observed in various fields, such as electrochemical dissolution of metals, Josephson junction, mercury beating heart, and flame of candles, irrespective of system details~\cite{Kiss2000,Cruz2007,Cruz2007a,Cruz2010,Wiesenfeld1996,Verma2014,Kumar2015,Biswas2017,Pantaleone2002,Okamoto2016,Kohira2012,Sharma2020,Heath1998,Cawthorne1999,Motter2013,Roy1994,Ashwin1998,Winful1992,Li1993}. In addition, striking synchronization phenomena appear in biological systems, such as fireflies, frog choruses, hair cells, heart cells, and so on~\cite{Glass2001,Smith1935,Buck1968,Moiseff2010,Jackson2019,Clay1979,Michaels1987,Neda2000,Walker1969,Reppert2002,Harrison2001,Takamatsu2000}. Such complex living systems are difficult to describe; however, the same simplicity is observed in their mechanisms.

The elevator cluster motion can be regarded as a synchronization as the following discussion demonstrates. First, elevators can be regarded as autonomous oscillators for many passengers~\cite{Tanida2021}. The elevator motion does not directly depend on time; it depends on the presence or absence of passengers, indicating autonomy. Moreover, the existing studies have proved that the inflow of passengers affects the cluster motion~\cite{Poschel1994,Hikihara1997,Nagatani2003,Nagatani2004,Nagatani2011,Nagatani2015,Nagatani2016,Feng2021,Tanida2021}; thus, the existence of passengers couples elevators with each other. 
However, it's not clear how much the elevators synchronize when elevators are not fully coupled. In this study, assuming that such a condition corresponds to the case that not all the passengers ride the first-arriving elevators, we introduced a control parameter, $\eta$, which is the proportion of passengers set to ride the first-arriving elevator. The proportion less than 1 indicates a situation, for example, where the distance between elevators is quite long for some passengers who have limited mobility. As an example that applies to this condition, we can assume elevators in hospitals whose main passengers are people in wheelchairs, elderly people, and patients. When the proportion is 0, the elevators are isolated. We numerically simulated behaviors of two elevators during peak loads when passengers go down to the ground floor to exit the building. We examined the order parameter of the elevator motion and round-trip time for various inflow rates of passengers and the proportion of passengers set to ride the first-arriving elevator. Subsequently, we demonstrate a self-consistent equation to calculate the round-trip time and a simple mathematical model to estimate the order parameter.

\section{Problem formulation}
\begin{figure}[h]
\centering
\includegraphics[width=.5\linewidth]{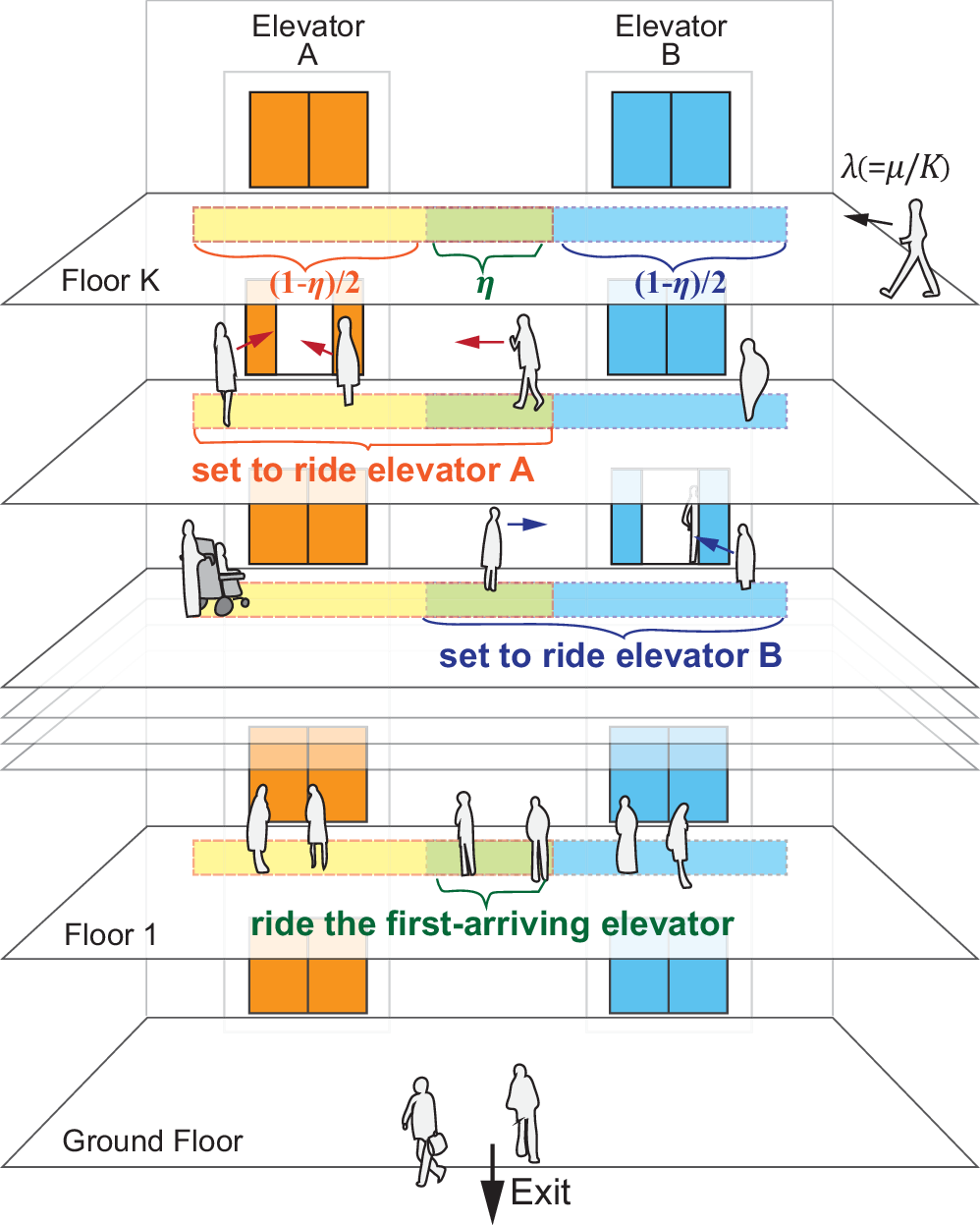}
\caption{
    Schematic of the model. We consider two elevators in a building with $K$ floors in addition to the ground floor. Passengers arrive in front of elevators on the upper floors following a Poisson process to move down to the ground floor. The proportion of passengers, $\eta$, can ride in an elevator that arrives earlier; Others can ride in only the predetermined elevator.
}
\label{fig:schematic}
\end{figure}

In this study, we considered a downward elevator system motion during peak loads. The elevator system consists of elevators A and B that serve $K$ floors (Fig.~\ref{fig:schematic}). The elevators use one time step to move one floor up or down and $\gamma$ time steps for the passengers to enter or exit. We set the coefficients of the model as $\gamma = 10$. Consistent with the existing studies \cite{Poschel1994,Hikihara1997,Tanida2021}, we assumed that all calls for elevators are from passengers waiting for the elevators on the $k$-th floor ($1 \leq k \leq K$) to move to the ground floor ($k=0$) and exit the building. Once an elevator goes down, it does not go up again until it arrives on the ground floor. If passengers exist neither in the elevator nor floor, both elevators stay on the ground floor until the next call. In the case of no-waiting calls and no passengers, the next call is accepted by the closest elevator to the calling floor. If both elevators are staying on the same floor without any passengers and a single new call occurs, one of the elevators is randomly selected to move. When more than one unresolved call exists and both elevators are free, the elevator that stopped on a higher floor (the upper elevator) moves upward. Simultaneously, the lower elevator starts moving if there are other calls it can reach faster compared to the upper one.
To estimate the time to go to the target floor $k_t$ from the floor $k$, the following additional assumptions are made: 1)	If the elevator is going up or on the ground floor, it will stop on the highest floor among those having unresolved calls, $k_h$, and then will switch it's direction of motion to down and stop on all floors with unresolved calls between $k_h$ and $k_t$. 2) If the elevator is going down and $k>k_t$, it will stop on all floors with unsolved calls between $k$ and $k_t$. 3) If the elevator is going down and $k_t \leq k$, it will stop on all floors with unresolved calls between $k$ and the ground floor and between $k_h$ and $k_t$. In addition, the estimated time is used to decide whether the lower elevator goes up or stops on a floor with unresolved calls, while the upper elevator is going down with passengers. Moreover, the actual time to arrive at the target floor is different from the estimated one because the number of waiting passengers is updated at every time step.

We assumed that the arrival of new passengers on each floor follows the Poisson law. Moreover, the Poisson parameter of each floor is uniform. The number of new passengers at each floor and every time steps $n$ is distributed according to the Poisson law:
\begin{eqnarray}
	P_{\lambda}(n) = \frac{\lambda^n}{n!}e^{-\lambda} \ ,
	\label{eq:poisson}
\end{eqnarray}
where $\lambda=\mu/K$ is the Poisson parameter, and $\mu$ represents the passenger inflow rate for the entire building.

In this study, we introduce a new parameter $\eta$. $\eta$ is the proportion of passengers who are set to ride the first-arriving elevator; Other passengers are set to ride only a predetermined elevator. Thus, the proportion of passengers set to ride only elevator A or B is $(1-\eta)/2$ each. The capacity of elevators is set to be large enough to accommodate any number of passengers based on the results of a previous study that proved elevators synchronize even if the capacity is extremely large~\cite{Tanida2021}.
As an initial condition, we set two elevators at random floor numbers and zero passengers on all floors. We simulated each parameter set for 11,000 time steps and used data after 1,000 time steps to discuss and analyze the mean values of the steady state. Finally, the elevators are assumed not to be sufficiently smart to identify the number of carrying passengers and possibly stop even if their capacity is full.

\section{Results}
\subsection{Simulation results of the order parameter and the round-trip time}
\begin{figure}[t]
\centering
\includegraphics[width=.99\linewidth]{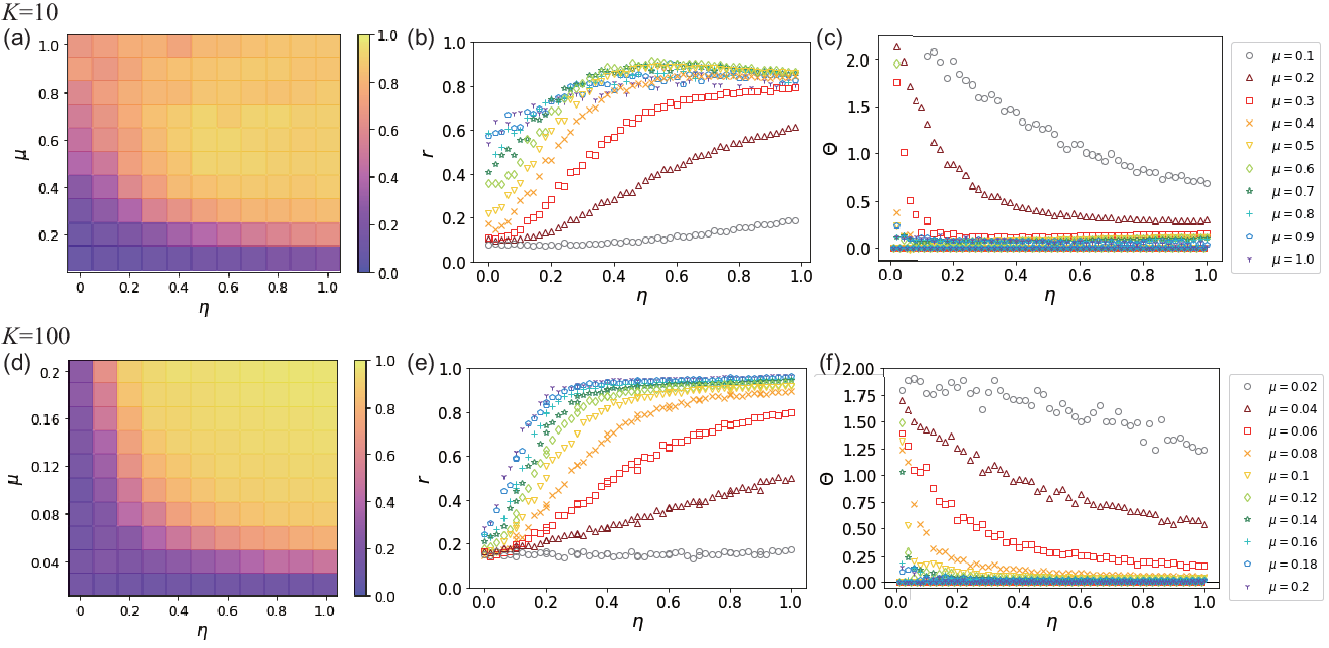}
\caption{
    [(a), (b), (d), and (e)]
    Values of the order parameter $r$ for various $\mu$ and $\eta$ at $K = 10$ and 100. In the heatmaps (a,d), the values of $r$ are indicated by the color. 
    [(c) and (f)] Values of the order parameter $\Theta$ for various $\mu$ and $\eta$ at $K = 10$ and 100.
}
\label{fig:order}
\end{figure}
First, we numerically simulated the model for various $\eta$, $\mu$, and $K$, and examined the order parameters $r$ and $\phi$, which is defined as
\begin{eqnarray}
	r e^{i\phi} = \frac{1}{N}\sum^{N}_{j=1} \exp \left(2\pi i \frac{ \tau_{j}}{T}\right) \ ,
\end{eqnarray}
where $\tau_j$ is the time between the $j$-th and ($j+1$)-th departures of either elevator on the ground floor. If both elevators arrive at the exact same time, $\tau_j$=0. $T$ is a mean of the round-trip time, which is defined as the time from the departure to the next arrival of each elevator. $N$ is the total number of departures. $r=1$ and $\phi=0$ when both elevators arrive simultaneously, $r=1$ and $\phi=\pi$ when $\tau_j$ for all $j$ equal to $T/2$, and $r$ is 0 when $\tau_j$ is uniformly distributed. 

\begin{figure}[t]
\centering
\includegraphics[width=.8\linewidth]{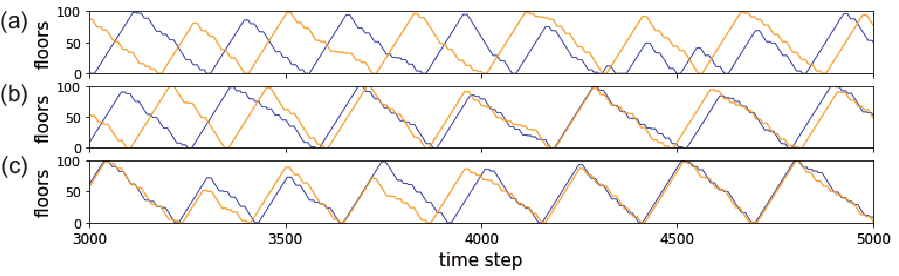}
\caption{
    Dynamics of the elevators for $\mu = 0.06$, $K = 100$, and different values of the passenger inflow rate: (a) $\eta = 0$, (b) $\eta = 0.3$, and (c) $\eta = 1.0$. The dark blue and bright orange lines represent the position of each elevator at a certain time step. As $\eta$ increases, the elevators tend to move as a group.
}
\label{fig:example}
\end{figure}

As shown in Figs.~\ref{fig:order}(a), \ref{fig:order}(b), \ref{fig:order}(d), and \ref{fig:order}(f), $r$ tends to increase with an increasing $\eta$. In particular, $r$ sharply increases at $\eta = 0$ and then stabilizes at $\eta = 0.3$ independent from $K$ for large $\mu$. In other words, the ordered and disordered states of motion generally emerge above and below around $\eta=0.3$, respectively.
When $r$ is close to 1, $\phi$ is close to 0 [Figs. (c) and (f)]. Figures~\ref{fig:example}(a-c) show examples of elevators' dynamics for various $\eta$ at $K=100$ and $\mu=0.06$. When $\eta=0$, two elevators move independently. When $\eta=0.3$ and 1, the positions of the two elevators tend to be close. Those results show that elevators exhibit in-phase synchronization when $\eta>0.3$.

Next, we examined the round-trip time $T$ for various $\eta$ and $\mu$ at $K=100$. As shown in Figs.~\ref{fig:roundtirptime}(a) and~\ref{fig:roundtirptime}(b), $T$ decreases with an increasing $\eta$, whereas it increases with an increasing $\mu$. To summarize the results so far, although both $\eta$ and $\mu$ are associated with promoting the emergence of synchronization, the effects of the two parameters on the round-trip time are different from each other; the round-trip time becomes shorter and longer as $\eta$ and $\mu$ increases, respectively.

\subsection{Numerical model for the round-trip time}
\begin{figure}[hbt]
\centering
\includegraphics[width=.8\linewidth]{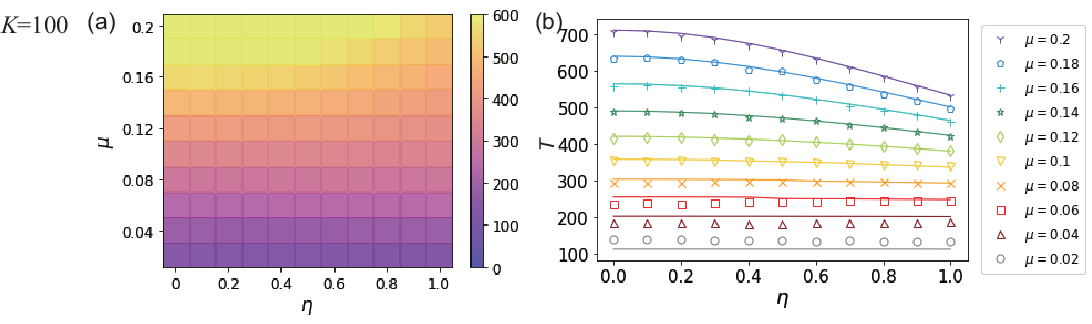}
\caption{
     Values of the round-trip time $T$ for various $\mu$ and $\eta$ at $K = 100$. The color of heatmap (a) represents the value of $T$. Solid lines represent theoretical values calculated using Eq.(\ref{eq:self}).
}
\label{fig:roundtirptime}
\end{figure}

To explain the relationship between the control parameters and outputs and clarify the mechanism of the cluster motion, we estimated the mean round-trip time for various $\mu$ and $\eta$ with a simple mathematical model. We assumed that $\mu$ is sufficiently large for elevators to move periodically, and the round-trip time $T$ is considerably larger than the stopping time step $\gamma$ for the passengers to enter or exit ($T \gg \gamma$). When elevator B arrives at a floor late by $\gamma$ compared to elevator A [Fig.~\ref{fig:theory}(a)], the average number of passengers waiting for elevator B at the floor in the duration $\gamma$ is $\mu \gamma/K$. Among those passengers, the proportion of $(1+\eta)/2$ can catch elevator B, whereas the proportion of $(1-\eta)/2$ needs to wait for elevator A [see also Fig.~\ref{fig:schematic}]. Moreover, the average number of passengers waiting on a floor where elevator A has just arrived is $\mu (T-\gamma)/K$; the proportion of $(1+\eta)/2$ can catch elevator A, and the proportion of $(1-\eta)/2$ needs to wait for elevator B. In this case, the probability that elevators A and B stop to serve are described as
\begin{eqnarray}
    P_A = 1- \exp\left(-\frac{\mu}{K} [(T-\gamma) \zeta + \gamma (1-\zeta)] \right)
	\ , \label{eq:Pa}\\
    P_B = 1- \exp\left(-\frac{\mu}{K} [(T-\gamma) (1-\zeta) + \gamma \zeta] \right)
	\ , \label{eq:Pb}
\end{eqnarray}
respectively, where $e$ is the probability that a passenger who cannot ride in one of the elevators, $\zeta=(1-\eta)/2$.
For $\eta>0$, $P_B$ is larger than $P_A$, indicating that both elevators become closer over time on average and eventually move together with the same timing. However, both elevators do not keep moving on exactly the same floor. One elevator might move forward while the other one might stop, which makes the closer time difference $\gamma$. Thus, we assumed that typical time differences are $\gamma$ and $T-\gamma$. The expectation value of the number of floors that an elevator serves on average can be described as 
\begin{eqnarray}
    n = K \frac{(P_A+P_B)}{2}
	\ .
\end{eqnarray}
In this model, we assumed that each elevator stops at $n$ floors. Subsequently, assuming that these $n$ floors are randomly distributed in the building, we estimated the typical highest floor in a round trip, ${\bar k}_h$, as follows:
\begin{eqnarray}
    {\bar k}_h 
    &=& \frac{1}{\Xi} \left\{
    \sum_{k=n+1}^K
    \left[
    \left( 
        \begin{array}{c}
            k \\
            n \\
        \end{array}
    \right) 
    -
    \left( 
        \begin{array}{c}
            k-1 \\
            n \\
        \end{array}
    \right)
    \right]k
    +
    \left( 
        \begin{array}{c}
            n \\
            n \\
        \end{array}
    \right)
    \right\}
    \nonumber \\
    &=& \frac{1}{\Xi} \left[
        \left( 
            \begin{array}{c}
                K \\
                n \\
            \end{array}
        \right)K
        -
        \sum_{\ell=0}^{K-n-1}
        \left( 
            \begin{array}{c}
                n+\ell \\
                n \\
            \end{array}
        \right) 
        \right]
    \label{eq:z}
	\ ,
\end{eqnarray}
where $ \Xi =
    \left(
    \begin{array}{c}
        K \\
        n \\
    \end{array}
    \right)$.
With $n$ and ${\bar k}_h$, a typical round-trip time $T$ can be described as a self-consistent function as follows:
\begin{eqnarray}
    T = 2{\bar k}_h + (n+1)\gamma
	\ . \label{eq:self}
\end{eqnarray}
Solid lines in Fig.~\ref{fig:roundtirptime}(b) show the value of $T$ obtained using Eq.(\ref{eq:self}). It can satisfactorily reproduce the typical round-trip time for various $\eta$ at each $\mu$, which are more appropriate than the results presented in a previous study~\cite{Tanida2021}. Thus, according to the exponent of the second term Eqs.(\ref{eq:Pa}) and (\ref{eq:Pb}), increasing $\eta$ (decreasing $\zeta$) is equal to reducing the round-trip time of each elevator.

\subsection{Numerical model for the order parameter}
\begin{figure}[t]
\centering
\includegraphics[width=.8\linewidth]{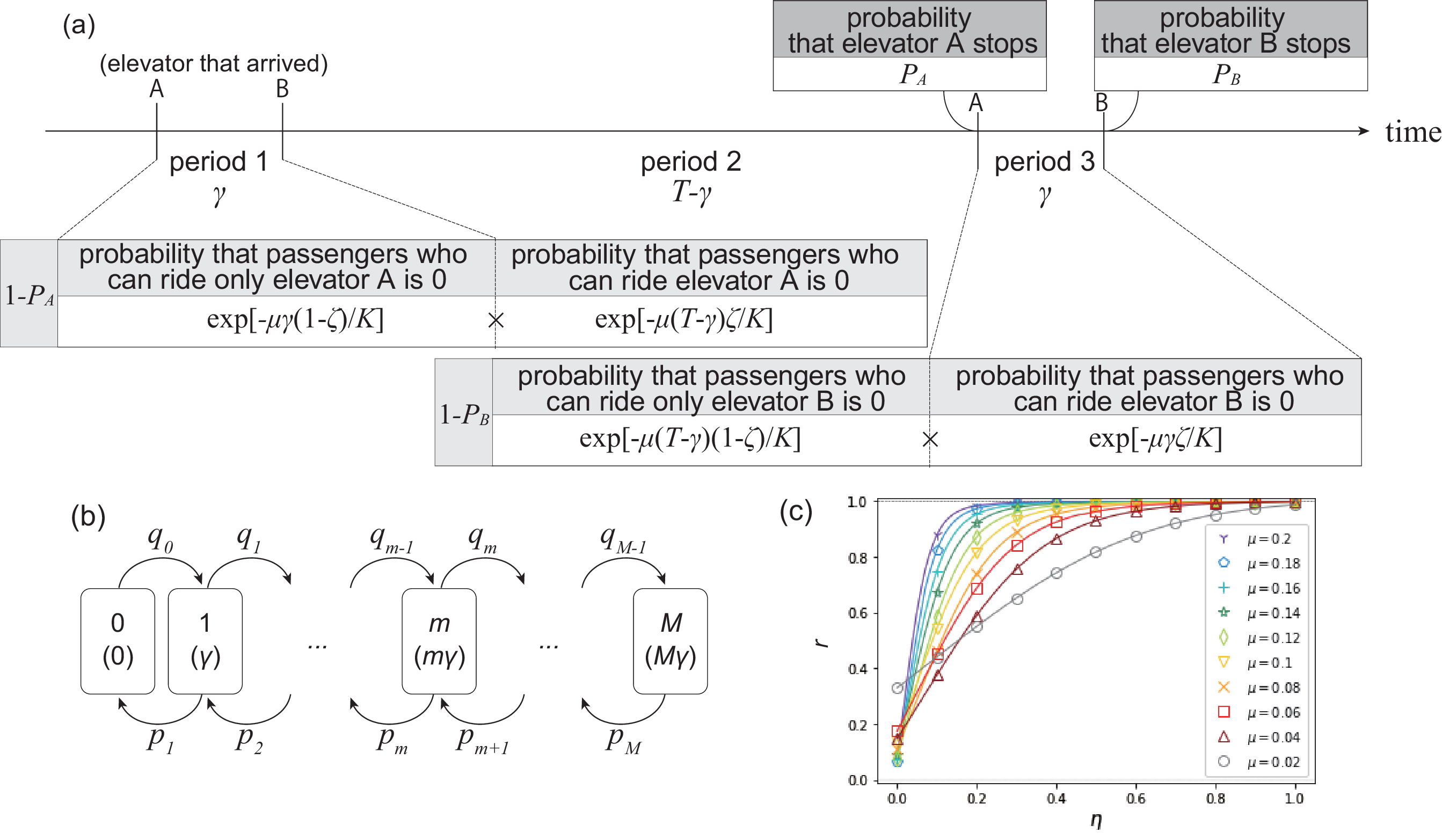}
\caption{
    a) Schematic of the relationship between the interval time of arrivals and timing to ride of passengers. (b) Schematic of the state transitions of the interval time. $q_m$ represents the probability to transit from state $m$ to $(m+1)$. $p_m$ represents the probability to transit from state $m$ to $(m-1)$. (c) Each marker with line represents the value of $r$ calculated using Eq.(\ref{eq:Smodel}) when $K = 100$ at each $\mu$.
}
\label{fig:theory}
\end{figure}
Based on Eq.(\ref{eq:self}), we roughly estimate the order parameter $r$ using a simple mathematical model in this section. We assumed that the round-trip time $T$ was constant. We defined the shorter time differences as $\tau$, where $\tau \leq T-\tau$. In addition, we assumed that $\tau$ gains a limited number of states; the value of $\tau$ is multiple times of $\gamma$, or $m\gamma$, where $m$ is an integer that satisfies $m\gamma \leq T/2$. We set $M$ as an integer, which satisfies $M=\lfloor T/\gamma \rfloor$. $\tau$ changes when only one of the elevators stops. We can interpret this event as the state transition from $m$ to $(m-1)$ or $(m+1)$ [Fig.~\ref{fig:theory}(b)]. For $m>0$, The probability of transition from $m$ to $(m-1)$ is represented by $p_m$. Additionally, the probability of state transition from $m$ to $(m+1)$ is represented by $q_m$ for $m\geq 0$. $p_m$ and $q_m$ are described as follows:
\begin{eqnarray}
p_m &=& P_A^{(m)}(1-P_B^{(m)}) \ (m>0) \\
q_m &=& \left\{
\begin{array}{ll}
(1-P_A^{(m)})P_B^{(m)} & (m> 0 )\\
(1-P_A^{(m)})P_A^{(m)} & (m= 0 ) \ ,
\end{array}
\right.
\end{eqnarray}
where $P_A^{(m)}$ and $P_B^{(m)}$ are the probabilities that elevators A and B stop, respectively, when we focus on the smaller time difference $m\gamma$. These probabilities are represented as follows:
\begin{eqnarray}
P_A^{(m)}&=& 1 - \exp\left(-\frac{\mu}{K} [(T-m\gamma)\zeta+ m\gamma (1-\zeta) ] \right) \\
P_B^{(m)}&=& 1 - \exp\left(-\frac{\mu}{K} [(T-m\gamma) (1-\zeta) + m\gamma \zeta ] \right) \ .
\end{eqnarray}
We represent the state probability at $m$, at which the time difference is $m\gamma$, as $x_m$.
The state probabilities can be calculated as follows;
\begin{eqnarray}
x_0^{(t+1)} &=& (1-q_0) x_0^{(t)} + p_1 x_1^{(t)} \ \mathrm{for}\ m=0\\
x_m^{(t+1)} &=& (1-p_m-q_m)x_m^{(t)} + q_{m-1}x_{m-1}^{(t)} + p_{m+1}x_{m+1}^{(t)} \nonumber \\
 &&\ \mathrm{ for}\ 0<m\leq\lfloor T/\gamma \rfloor \ ,
\end{eqnarray}
which satisfy the normalized condition: $\sum_{m=0}^{M} x_m = 1$. Since we have made the ergodic assumption, the steady state is uniquely determined. Because the order parameter $r$ at $m$ is $\cos(m {2\pi\gamma}/{T})$, the order parameter r averaged for a long time is described as follows: 
\begin{eqnarray}
r = \sum_0^{M} x_m \cos\left(m \frac{2\pi\gamma}{T}\right)
\ . \label{eq:Smodel}
\end{eqnarray}
Figure~\ref{fig:theory}(c) demonstrates $r$ calculated using this simple model. For $\mu$ smaller than 0.1, $r$ becomes larger than that of simulation results. It could be because that round-trip time would not be constant. Nevertheless, in general, $r$ as a function of $\eta$ satisfactorily agrees with that of the simulation for a large $\mu$. $r$ rapidly increases at small $\eta$ and saturates at approximately $\eta = 0.3$, which agrees with the simulation results in Figs.~\ref{fig:order}(c) and \ref{fig:order}(d).

\section{Discussion and Conclusions}
In this study, we investigated differences between the dynamics of elevators when they were isolated or coupled by introducing a control parameter, $\eta$, the proportion of passengers who are set to ride the first-arriving elevators. Considering elevators’ downward motion during peak loads when passengers go down to the ground floor to exit the building, we simulated the elevators' dynamics for various $\eta$ and the inflow of passengers $\mu$, and calculated the order parameters and the round-trip time. The order parameters generally increased as $\eta$ or $\mu$ increased. They rapidly increased at small values of $\eta$ and became saturated at approximately $\eta = 0.3$ with a sufficiently large $\mu$. The round-trip time decreased with an increasing $\eta$, whereas it increased as $\mu$ increased.

To discuss essential differences in the effects of $\mu$ and $\eta$ on the round-trip time, we established a simple mathematical model for the round-trip time as a function of $\mu$ and $\eta$. In this model, increasing $\mu$ promotes the probability to stop any elevators independent of the arrival time, whereas an increasing $\eta$ decreases the probability to stop the one that comes later of the two elevators moving almost simultaneously. The model satisfactorily reproduced the simulation results of the round-trip time. Furthermore, based on the model, we introduced a mathematical model to explain the behavior of $r$ as a function of $\mu$ and $\eta$. Indeed, it can capture a comparable trend as that of $r$ in numerical simulations.

Generally, A high building often has more than two elevators, and the conformation of the elevator hall is varied for each building. Some have elevators aligned on the same side, others have elevators facing each other. In those cases, the relations between each elevator are not uniform. The relation of elevators can be varied among each pair depending on the distance, the side that elevators are facing, and other factors. Passengers on wheelchairs, for example, could ride the adjacent elevators, while they might decide not to ride on the first-arriving elevators if it is too far. To represent such various relationships of elevators in the same elevator hall, we believe parameter $\eta$ is useful.

\section*{Acknowledgment}
We are grateful to Katsuhiro Nishinari, Daichi Yanagisawa, Yuki Koyano, Kaori Sugimura, Jia Xiaolu, and Claudio Feliciani for helpful discussions and for their kind interest in this work.


\end{document}